\documentstyle[12pt]{article}

\newcommand{\nc}{\newcommand}
\nc{\be}{\begin{equation}}
\nc{\ee}{\end{equation}}
\nc{\bea}{\begin{eqnarray}}
\nc{\eea}{\end{eqnarray}}
\nc{\bd}{\begin{displaymath}}
\nc{\ed}{\end{displaymath}}
\nc{\s}{\sum}
\nc{\vi}{Virasoro}
\nc{\wh}{\widehat}
\nc{\ra}{\rightarrow}
\nc{\pa}{\partial}
\nc{\dv}{$D$--Virasoro}
\nc{\dvr}{$D$-Vir}
\nc{\sdv}{$D$-Vir$^{(+,0)}$}
\nc{\al}{\alpha}
\nc{\til}[1]{\tilde{#1}}
\nc{\lb}[1]{\label{eqn:#1}}
\nc{\rf}[1]{\ref{eqn:#1}}
\nc{\lag}{\langle}
\nc{\rag}{\rangle}
\nc{\hf}{1\over2}
\nc{\qint}{{{q^{x}-q^{-x}}\over{q-q^{-1}}}}
\nc{\ze}{\zeta}

\begin{document}

\begin{titlepage}
\rightline{TMUP-HEL-9407}
\rightline{November 1994}

\vskip 1.5cm

\begin{center}
\begin{Large}
{  $W_{1+\infty}$ as a Discretization of Virasoro Algebra}
\end{Large}

\vskip 4cm

Ryuji KEMMOKU\footnote[2]{e-mail:\ kemmoku@phys.metro-u.ac.jp} and Satoru
SAITO\footnote[3]{e-mail:\ saito@phys.metro-u.ac.jp}
\vskip  .5cm

{\it Department of Physics, Tokyo Metropolitan University}

{\it Minami-Osawa, Hachioji, Tokyo 192-03, Japan}

\end{center}

\vskip 8cm
\centerline{\bf Abstract}

\vskip 0.5cm
It is shown that the $W_{1+\infty}$ algebra is nothing but
 the simplest subalgebra of a $q$-discretized \vi\ algebra
(\dv),
in the language of the KP hierarchy explicitly.
\vskip .5cm

\end{titlepage}
\newpage

\section{Introduction}

In the research of integrable systems,
discretization of variables provides important information about the systems.
Stability under discretization characterizes a certain class of integrable
systems which is completely integrable.
For example every soliton equation of the KP hierarchy \cite{sat,djkm} has
a discrete analog
which is common to all equations.
Namely a single bilinear difference equation of Hirota
reproduces every soliton equation of the KP
hierarchy by taking certain continuous limit of variables \cite{hiro}.
This remarkable property of soliton type equations should be compared with a
generic case in which discretization of variables in general make nonlinear
equations create chaos \cite{ns}.
This means that there exist some symmetries which preserve
integrability under the discretization of variables.

Discretization of differential operators also plays a key role
 in the $q$-deformed conformal field theories.
The $q$-deformed Knizhnik-Zamolodchikov ($q$-KZ) equation
by Frenkel and Reshetikhin \cite{f-r} has been
formulated by
the consideration of
representation theory of $U_{q}(\widehat{sl_2})$,
and is essentially realized through discretization of variables of
the original KZ equation.

We are interested in relations which remain true irrespective to continuous
or discrete.
Such relations will not only characterize completely integrable systems but
also cast a light to understand the boundary
between deterministic and nondeterministic nature of nonlinear equations.

In order to claim this idea we proposed, in a series of papers \cite{kem},
a deformation of the
Virasoro algebra which was realized by a $q$-discretization of differential
 operators and called it \dv.
This algebra was shown to admit free field representation both of fermionic
and bosonic \cite{chai}.
It was constructed such that the \vi\ algebra was reproduced
in the continuous limit ($q\ra1$).
Although it was derived as general as possible, there
remains a problem to clarify the relation to other known symmetries.

In this paper
we show explicitly that the simplest subalgebra of \dv\ is nothing but
the $W_{1+\infty}$ algebra \cite{pop}.
This result provides a new interpretation of the $W_{1+\infty}$ algebra.
Namely it enables us to understand that the $W_{1+\infty}$ algebra emerges
as a result of a proper discretization of the \vi\ algebra.

This paper is organized as follows:
In Section 2, we review on the discretization of the \vi\ algebra.
In Section 3, we apply it to the KP hierarchy.
For the purpose of it, we first review on additional symmetry of the
KP hierarchy which is found in \cite{chen} and generalized in \cite{orl}.
Next we investigate how the flow on the
universal Grassmann manifold induced by
$q$-shift operators can be expressed in terms of the symmetry, and denote
it by $E_{mn}$.
Based on the results of \cite{ow}, we show that the action of $E_{mn}$
on the $\tau$ function can be realized
by use of the $q$-shift operators and the vertex operators of the KP hierarchy.
In the last part, we construct operators by a proper combination
of $E_{mn}$.
They will be shown equivalent to generators of the simplest \dv\
subalgebra as the action on the $\tau$ function.
Finally, by using the fact that the vertex operators give generators of the
$W_{1+\infty}$ algebra \cite{dic2,tt},
we find that the subalgebra is the $W_{1+\infty}$ algebra.

\section{A brief review on \dv}

In \cite{kem}, we proposed the following algebra (without central extension);
\bea
\Bigl[ {\cal L}_{m}^{(n,r;\pm)}, {\cal L}_{m'}^{(n',r;\pm)} \Bigr]
&=&
C(_{m'\ n'+ r}^{m\ \ n+r})_{\pm}\ {\cal L}_{m+m'}^{(n+n'+r,r;\pm)}+
C(_{m'\ n'-r}^{m\ \ n-r})_{\pm}\ {\cal L}_{m+m'}^{(n+n'-r,r;\pm)}
\nonumber \\
&& \hspace{.1in} + C(_{m'\ -n'+r}^{m\ \ \  n+r})_{\pm}
\ {\cal L}_{m+m'}^{(n-n'+r,r;\pm)}
+C(_{m'\ -n'-r}^{m\ \ \ n-r})_{\pm}\ {\cal L}_{m+m'}^{(n-n'-r,r;\pm)}
\lb{dv2}
\eea
for each value of $r$.
The double signs on both sides correspond each other.
The structure constant $C$ is given by
\be
C(_{m'\ n'+r}^{m\ \ n+r})_{\pm} \equiv
{{\Bigl[ {{(n+r)m'-(n'+r)m} \over 2} \Bigr]_{-}\  [n+n'+r]_{\mp}} \over
{2\  [n]_{\mp}\  [n']_{\mp}\  [r]_{\pm}}}\ ,
\ee
where
\bd
 [x]_{+}\equiv{{q^{x}+q^{-x}} \over 2}\ ,\
 [x]_{-}\equiv\qint\ .
\ed
The generators ${\cal L}_{m}^{(n,r;\pm)}$ are realized as
\be
{\cal L}_{m}^{(n,r;\pm)}= z^{m}
{\left[ n \left(z\pa_{z} +{m \over 2} \right) \right] _{\mp} \over [n]_{\mp}}
         { \Bigl[ r z\pa_{z} \Bigr]_{\pm} \over [r]_{\pm}}
  \equiv  z^{m}
\left[ z\pa_{z} +{m \over 2}  \right] _{n;\mp}
         \Bigl[ z\pa_{z} \Bigr]_{r;\pm} \ .
\lb{dvi}
\ee
In (\rf{dvi}), ${\cal L}_{m}^{(n,r;\pm)}$ essentially consist of
$q$-difference operators.
Moreover, in the limit of $q\ra1$, ${\cal L}_{m}^{(n,r;\pm)}$ reduce to the
Virasoro
generators and (\rf{dv2}) becomes the commutation relation of the Virasoro
algebra.
Then we named it \dv\ algebra (\dvr).

We also have the central extended version of this algebra which are
represented by free fields as follows;
\be
{\hat{\cal L}}_{m}^{(n,r;\pm)}={1 \over 2}\s_{k} A_{k,m-k}^{(n;\pm)} (q)
\ :\al_{m-k}^{(r;\pm)}\   \al_{k}^{(r;\pm)}:.
\lb{bil}
\ee
where $\al_{k}^{(r;\pm)}$ satisfy the following relations;
\be
{[\al_{k}^{(r;\pm)},\al_{k'}^{(r;\pm)}]}_{\pm}\equiv
\al_{k}^{(r;\pm)}\al_{k'}^{(r;\pm)}\pm\al_{k'}^{(r;\pm)}\al_{k}^{(r;\pm)}
=D_{k}^{(r;\pm)}(q)\
\delta_{k+k',0}\ ,
\lb{com}
\ee
i.e. $\al_{k}^{(r;+)}$ denotes `fermion',
and $\al_{k}^{(r;-)}$ does `boson', respectively.
We get simple solutions for $A$ and $D$;
\be
A_{k,m-k}^{(n;\pm)}=-\biggl[ {{2k-m} \over 2} \biggr]_{n;\mp}\ , \
D_{k}^{(r;\pm)}=[ k ]_{r;\pm}\ .
\lb{36}
\ee

In the following discussion, we will not use the full structure of this
algebra.
As our attention is paid to its relation to the KP hierarchy,
we only use a subalgebra $D$--Vir$^{(+,r=0)}$ of which generators
$\{{\cal L}_{m}^{(n,0;+)}\}\ (n\in{\bf Z}_{>0})$ are realized by use of the
ordinary fermion like as the KP hierarchy.

\section{Action of \dv\ to the KP hierarchy}

\subsection{Additional symmetries of the KP hierarchy}
Let us start with the KP hierarchy with Lax representation \cite{sat,djkm};
\bea
&&\pa_{l}L=[L_{+}^{l},L]
\lb{lax} \\
&&L=W\pa W^{-1},\hspace{.3in}
W=1+w_{1}(x)\pa^{-1}+w_{2}(x)\pa^{-2}+\cdots, \nonumber
\eea
where $L_{+}$ denotes the differential operator part of the pseudo-differential
operator $L$, then $L=L_{+}+L_{-}$, and
$\pa=\pa/\pa x,\ \pa_{l}=\pa/\pa t_{l}\ (l\in{\bf Z}_{>0},\ x=t_{1})$.
Any equation of (\rf{lax}) generates symmetry for other equations, and they
are commutative in the sense that $[\pa_{l},\pa_{l'}]=0$.
By removing the dressing, (\rf{lax}) becomes $[\pa_{l}-\pa^{l},\pa]=0$.

In addition to this symmetry it is known that other symmetries exist
\cite{chen,orl}.
First we introduce an operator
\be
\Gamma=\s_{r=1}^{\infty}rt_{r}\pa^{r-1}.
\ee
It is clear that $[\pa_{l}-\pa^{l},\Gamma]=0$.
Then if we define $M=W\Gamma W^{-1}$, we can easily check
$[\pa_{l}-L_{+}^{l},M]=0$.
(The essential difference between $M$ and $L$ is in the expansion coefficients.
The coefficients of $M$ may depend on the KP time $\{t_{l}\}$ explicitly.)
By combining this equation with (\rf{lax}) and generalizing them, we get
\be
[\pa_{l}-L_{+}^{l}, M^{k}L^{m}]=0 \hspace{.5in}\forall\ k\in{\bf Z}_{>0},
m\in{\bf Z}
\ee
If we introduce new variables $t_{mk}$ satisfying the following equation as
\be
\pa_{mk}L=-[(M^{k}L^{m})_{-},L] \hspace{.5in}\pa_{mk}={\pa\over{\pa t_{mk}}},
\lb{4}
\ee
or in terms of $W$ operators;
\be
\pa_{mk}W=-(M^{k}L^{m})_{-}W,
\lb{new}
\ee
we can prove that $\{\pa_{mk}\}$ commutes with the KP flow, i.e.
$[\pa_{mk},\pa_{l}]=0$.
In this sense the flow $\{\pa_{mk}\}$ were called
{\it additional} symmetries of the KP hierarchy\footnote[2]{If
$k=0$, $\pa_{m0}$ is $\pa_{m}$ for $m>0$.} \cite{dic2}.
But the new flow themselves do not commute with each other:
Since the operators $L$ and $M$ are canonically conjugate, i.e. $[L,M]=1$,
 we can
consider the homomorphism such as $L\mapsto\pa$ and $M\mapsto x$.
This mapping enables us to calculate the commutation relation of
$\{\pa_{mk}\}$.
Actually, we get
\be
[\pa_{mk},\pa_{m'k'}]
=\s_{j=1}^{\infty}\Biggl\{
\left(
\begin{array}{c}
m\\
j
\end{array}
\right)
\left(
\begin{array}{c}
k\\
j
\end{array}
\right)
-
\left(
\begin{array}{c}
m'\\
j
\end{array}
\right)
\left(
\begin{array}{c}
k'\\
j
\end{array}
\right)
\Biggr\}\ j!\ \pa_{m+m'-j,k+k'-j}.
\lb{mk}
\ee
In a simple case as $\{\pa_{m+1,1}\}$,  it is well-known that they
form an algebra isomorphic to \vi\ algebra (without central extension).
Remark that (\rf{4}) and (\rf{new}) have the degree of freedom for the gauge
choice.
But we will fix the gauge in the above form in the following discussion.


Now we show the action of the new symmetry to an element of the universal
Grassmann manifold (UGM).
Let $H$ be the space of formal expansions $H=\{f(z)=\s f_{k}z^{k}\}$.
If a subspace $V\subset H$ has a natural bijection $V\ra H_{+}$,
 where $H_+$ is the positive power part of $H$,
then UGM is defined by
UGM\ =\ $\{V\subset H | V\simeq H_{+}\}$.
It is known that there is a {\it monic} $z$-operator\footnote[3]
{In a standard form, $z$-operators are defined by
\bd
G=G(\pa_{z},z)=
\s_{i\leq0,j\geq0}a_{ij}z^{i}\pa_{z}^{j}.
\ed}
$G=1+\s a_{ij}z^{i}\pa_{z}^{j}$ such that $V=GH_{+}$.
So generically, for $v\in V$ and $h_{+}\in H_{+}$, there is a monic operator
$W(t^{*},\pa_{z},z)$ such as $e^{-\xi^*}v=W(t^{*},\pa_{z},z)h_{+}$ where
$\xi^{*}=\xi(t^{*},z)=\s_{r=2}^{\infty} t_{r} z^{r}$.
If we consider the replacement as $z\mapsto\pa$, $\pa_{z}\mapsto x$ and write
the factors of the operators in an inverse order, we see that $z$-operators and
pseudo-differential
operators are related as an anti-isomorphism;
$G(\pa_{z},z)\mapsto G(x,\pa)=\s a_{ij}x^{j}\pa^{i}$.
By using this formula, we can prove that
$W(t^{*},x,\pa)$ is nothing the dressing operator of the KP hierarchy.
 (see \cite{dick} for more detail)

We can consider that the additional symmetry flow on UGM
is given by the operators $\pa_{mk}:V\ra H/V$.
Namely
\bea
\pa_{mk}\  e^{-\xi^{*}}v
&=&\pa_{mk}\ W(\pa_{z},z)h_{+}\nonumber \\
&&\hspace{-.5in}=-W\biggl(L(\pa_{z},z)^{m}M(\pa_{z},z)^{k}\biggr)h_{+}
  \hspace{.3in}({\rm mod}\ V) \nonumber \\
&&\hspace{-.5in}=-z^{m} \biggl(\pa_z +\s_{2}^{\infty}rt_{r}z^{r-1}\biggr)^{k}\
e^{-\xi^{*}}v =- e^{-\xi^*}z^{m} \pa_{z}^{k}\ v. \nonumber
\eea
The last equality can be obtained by induction.
If we use $[\pa_{mk},\pa_{i}]=0$, the l.h.s. becomes
$\pa_{mk}\  e^{-\xi^{*}}v= e^{-\xi^{*}}\pa_{mk}\ v$.
Then dividing by $e^{-\xi^{*}}$ on both sides, we get
\be
\pa_{mk}\ v=-z^{m} \pa_{z}^{k}\ v.
\lb{v}
\ee
In general, the algebra which consists of the operator $\{z^{m}\pa_{z}^{k}\}$
is called the $W$-infinity algebra.
Then the flow $\pa_{mk}$ is nothing but the action of the
$W$-infinity on UGM.

\subsection{Flow induced by shift operators}

To apply the result of above discussion to our theory, we first expand
$q$-shift operators $q^{nz\pa_{z}}$ with respect to the differential operators,
\be
q^{nz\pa_{z}}=\s_{j=0}^{\infty}{{(n\lambda)^{j}}\over{j!}}(z\pa_{z})^{j}
=\s_{j=0}^{\infty}{{(n\lambda)^{j}}\over{j!}}\s_{k=0}^{\infty}c_{jk}
\ z^{k}\pa_{z}^{k}\ ,
\lb{exp}
\ee
where $\lambda=\ln q\ (q^{N}\neq 1;\ \forall N\in{\bf Z}_{>0})$
and the coefficients $c_{jk}$ is given by
\be
c_{jk}=\s_{\al=1}^{k}{{(-1)^{k-\al}\ \al^{j}}\over{(k-\al)!\ \al!}}
\hspace{.2in} (j,k\geq1),
\hspace{.3in} c_{l0}=c_{0l}=\delta_{l,0}\ .
\lb{cjk}
\ee
Remark that we can solve (\rf{exp}) conversely,
\be
z^{k}\pa_{z}^{k}=\s_{j=0}^{\infty}d_{kj}\ (z\pa_{z})^{j}
=\s_{j=0}^{\infty}d_{kj}\ j!\ n^{-j}\
 \oint {d\lambda\over 2\pi i}\  \lambda^{-j-1}\ e^{n\lambda z\pa_{z}},
\lb{cv}
\ee
where $d_{kj}$ is the inverse of the ($\infty\times\infty$) matrix
$c_{jk}$ and satisfy
the following relations;
\be
d_{kj}=(-1)^{k+j}\ \{d_{k-1,j-1}+(k-1)\ d_{k-1,j}\}
\hspace{.2in}(j,k\geq1),
\hspace{.3in}d_{l0}=d_{0l}=\delta_{l,0}.
\ee

Now we consider a form as
\be
E_{mn}\equiv
-q^{mn\over2}\s_{j=0}^{\infty}{{(n\lambda)}^{j}\over{j!}}
    \s_{k=0}^{\infty} c_{jk}\ \pa_{m+k,k}\ .
\ee
{}From (\rf{v}) and (\rf{exp}), we understand that
the flow $E_{mn}$ on UGM is equivalent to the action of $q$-shift
operator, that is
\be
E_{mn}\ v= z^{m} q^{n(z\pa_{z}+{m\over2})}\ v.
\lb{dmn}
\ee
So if we remember (\rf{dvi}), we can write the action of \sdv\ on UGM
in terms of additional symmetry flow;
\be
{\cal L}_{m}^{(n,0;+)}\ v=
\left[{{E_{mn}-E_{m,-n}}\over{q^{n}-q^{-n}}}\right]\ v.
\ee
In this form, we see that ${\cal L}_{m}^{(n,0;+)}$ is nothing but the
$q$-difference operator of the spectral parameter $z$ of the KP hierarchy.
It is easy to check that in the $q\ra1$ limit,
\bd
{\cal L}_{m}^{(n,0;+)}\ v=
z^{m}\left[ z\pa_{z} +{m \over 2}  \right] _{n;-} v\
\longrightarrow\  -\pa_{m+1,1}\ v.
\ed

Next we consider the action of $E_{mn}$ on $\tau$ function.
Based on the above discussion, we first write down the action of the operator
$z^{m}\pa_{z}^{k}$ on the wave function $w(z)=We^{\xi(t,z)}$ in terms of
pseudo-differential operators;
\be
z^{m}\pa_{z}^{k}w(z)=W(x,\pa)\Gamma^{k}\pa^{m}e^{\xi}.
\ee
After multiplying the adjoint wave function
$w^{*}(z)=(W^{*})^{-1}e^{-\xi(t,z)}$
from the right on both sides, we integrate along a contour around $z=\infty$;
\be
\oint {{dz}\over{2\pi i}}\ \Bigl(z^{m}\pa_{z}^{k}w(z)\Bigr)\ w^{*}(z)
=\oint {{dz}\over{2\pi i}}\ \Bigl(W(x,\pa)\Gamma^{k}\pa^{m}e^{xz}\Bigr)\
 \Bigl((W^{*})^{-1}e^{-xz}\Bigr),\nonumber\\
\lb{int}
\ee
and look at how each side is expressed by using the $\tau$ function.\\
$\circ$ {\it The l.h.s. of (\rf{int})}

The wave functions can be written as
\be
w(z,t)={V(z)\ \tau \over \tau},\hspace{.2in}
w^{*}(z,t)={V^{*}(z)\ \tau \over \tau},
\ee
where the vertex operators $V$ and $V^*$ are defined as
{
\setcounter{enumi}{\value{equation}}
\addtocounter{enumi}{1}
\setcounter{equation}{0}
\renewcommand{\theequation}{\theenumi\alph{equation}}
\bea
&&V(z)=\exp \left(\s_{r=1}^{\infty} z^{r} t_{r}\right)
    \  \exp \left(-\s_{r=1}^{\infty} {1\over r}z^{-r} \pa_{r}\right),\\
&&V^{*}(z)=\exp \left(-\s_{r=1}^{\infty} z^{r} t_{r}\right)
    \  \exp \left(\s_{r=1}^{\infty} {1\over r}z^{-r} \pa_{r}\right),
\eea
\setcounter{equation}{\value{enumi}}
}
respectively.
They satisfy the following anti-commutation relation as
\be
\{V(z),V^{*}(\ze)\}=\delta\biggl({z\over\ze}\biggr).
\lb{ant}
\ee
The $\delta$
function is defined by formal expansions as
\be
\delta(z)=\s_{k\in{\bf Z}}z^{k}.
\ee
If we use (\rf{ant}) and the bilinear identity for the wave functions;
\be
\oint{dz \over 2\pi i}\ w(z,t)\ w^{*}(z,t')=0,
\ee
we can write the bilinear form  $w(z)w^{*}(\ze)$ by means of the vertex
operators
and the $\tau$ function as
\bd
V(z)V^{*}(\ze)\ (-\pa\ln \tau).
\ed
Hence the l.h.s. of (\rf{int}) is written as
\be
\oint {{dz}\over{2\pi i}}\ \Bigl(z^{m}\pa_{z}^{k}V(z)\Bigr)\ V^{*}(z)\
(-\pa\ln\tau).
\lb{vtx}
\ee
$\circ$ {\it The r.h.s. of (\rf{int})}

We use the following lemma \cite{ow,dic2};

\vspace{.2cm}

\noindent{\bf Lemma}\hspace{.1in}
{\it For any two pseudo-differential operators $P$ and $Q$},
\be
\oint {dz\over{2\pi i}}\ (Pe^{zx})(Qe^{-zx})={\rm res}_{\pa}\ PQ^{*},
\ee
{\it where}\ \ ${\rm res}_{\pa}\s a_{k}\pa^{k}=a_{-1}$.\ \ $\Box$

\vspace{.2cm}

\noindent Then the r.h.s. of (\rf{int}) becomes
\bea
&&{\rm res}_{\pa}\ W\Gamma^{k}\pa^{m}W^{-1}=
{\rm res}_{\pa}\ M^{k}L^{m}
={\rm res}_{\pa}\ (M^{k}L^{m})_{-}W\nonumber\\
&&\hspace{.5in}=-{\rm res}_{\pa}\ \pa_{mk}w(z)
 =-\pa_{mk}w_{1}=\pa_{mk}(\pa\ln \tau).
\lb{wave}
\eea

{}From (\rf{vtx}) and (\rf{wave}), we find that the flow $\pa_{mk}$
induced by the action of $z^{m}\pa_{z}^{k}$ on UGM acts on the $\tau$
function in the following way \cite{ow};
\be
\pa_{mk}\ \tau=-\left[\oint {dz\over{2\pi i}}\ z^{m} {{\pa^{k}V(z)}\over{\pa
z^{k}}}  V^{*}(z)\right]\ \tau.
\lb{25}
\ee
This expression enables us to write the flow $E_{mn}$ on the $\tau$ function
in more familiar form.
Actually, by using (\rf{dmn}) and (\rf{25}) we obtain
\bea
E_{mn}\ \tau
&=&\left[\oint {dz\over{2\pi i}}\ z^{m}\Bigl( q^{n(z\pa_{z}+{m\over2})}
\ V(z)\Bigr)
            \   V^{*}(z)\right]\ \tau \nonumber \\
&=&\left[{q^{{nm\over2}}\over{1-q^{-n}}}\oint {dz\over{2\pi i}}\ z^{m}X(q^{n}z,
z)\right] \ \tau.
\lb{shift}
\eea
where the vertex operator $X(z,\ze)\equiv:V(z)V^{*}(\ze):$\ .

\newpage
\subsection{$W_{1+\infty}$ as \sdv}

The final result of the preceding part is very important for our discussions:
In the theories of the KP hierarchy, the vertex operators relate to
the generators of the $W_{1+\infty}$ algebra.
Then if we can show that the transformation $\tau\ra\tau+E_{mn}\tau$
relates to \dv,
we understand the relation between $W_{1+\infty}$ and \dv.
So we first investigate how the flow $E_{mn}$ connects with \dv.

We consider the Fock representation of the KP hierarchy \cite{djkm}.
Let $\psi_{l}$ and $\psi^{*}_{l}$ ($l\in{\bf Z}+{\hf}$) be free fermions
such as $\{\psi_{l}, \psi^{*}_{l'} \} =\delta_{l+l',0}$
and $\{\psi_{l}, \psi_{l'} \}
=\{\psi^{*}_{l}, \psi^{*}_{l'} \} =0$.
The vacuum $\lag0|$ and $|0\rag$ are defined by
{
\setcounter{enumi}{\value{equation}}
\addtocounter{enumi}{1}
\setcounter{equation}{0}
\renewcommand{\theequation}{\theenumi\alph{equation}}
\bea
&&\psi_{l}|0\rag=0\ (l<0),\hspace{.3in}\psi^{*}_{l}|0\rag=0\ (l>0),\\
&&\lag 0|\psi_{l}=0\ (l>0),\hspace{.3in}\lag 0|\psi^{*}_{l}=0\ (l<0).
\eea
\setcounter{equation}{\value{enumi}}
}
In this representation, the $\tau$ function can be written as
\be
\tau(t,g)=\lag 0| e^{H(t)}g|0\rag \hspace{.3in}g\in GL(\infty)
\ee
where
\bd
H(t)=\s_{r=1}^{\infty}H_{r}t_{r},\ \  H_{r}=\s_{l\in{\bf Z}+{\hf}}
:\psi_{l}\psi^{*}_{r-l}: ,
\ed
or equivalently
\bea
&&\hspace{2cm}:\psi(z)\psi^{*}(z):=\s_{r\in {\bf Z}}H_{r}z^{-r-1},\nonumber\\
&&\psi(z)=\s_{l\in{\bf Z}+{\hf}} \psi_{l}z^{-l-{\hf}},\hspace{.3in}
\psi^{*}(z)=\s_{l\in{\bf Z}+{\hf}} \psi^{*}_{l}z^{-l-{\hf}}.\nonumber
\eea
The action of $:\psi(z)\psi^{*}(\ze):$ on the (neutral) state
$g|0\rag$ corresponds to the action of the operator
\be
Y(z,\ze)\equiv{1\over{z-\ze}}(X(z,\ze)-1)
\ee
on $\tau(t,g)$, that is
\be
Y(z,\ze)\ \tau(t,g)=\lag 0|e^{H(t)}:\psi(z)\psi^{*}(\ze):g|0\rag.
\ee
If we set a form as
\be
L_{m}^{(n)}={1\over{q^{n}-q^{-n}}}\ \left\{
q^{-n}E_{m-1,n}-q^{n}E_{m-1,-n}-
{{\cosh( \lambda n(m-1)/2)}\over{\sinh (\lambda n/2)}}\ \delta_{m,1}\right\},
\lb{tom}
\ee
the action of $L_{m}^{(n)}$ to the $\tau$ function becomes
\bea
L_{m}^{(n)}\ \tau&=&
\oint{dz\over 2\pi i}\ \lag 0|e^{H(t)}:\Bigl({\cal L}_{m}^{(n,0;+)}\psi(z)
\Bigr)\ \psi^{*}(z):g|0\rag\nonumber \\
&=&
\lag 0|e^{H(t)}\biggl(-{\hf} \s_{l}\biggl[ {{2l-m+1} \over 2} \biggr]_{n;-}
:\psi_{l}\psi^{*}_{m-l}:\biggr)g|0\rag \nonumber\\
&=&\lag 0|e^{H(t)}\hat{\cal L}_{m}^{(n,0;+)}g|0\rag,
\eea
where $\hat{\cal L}_{m}^{(n,0;+)}$ is given by (\rf{bil}).
We can easily check the following identity;
\be
\Bigl[L_{m}^{(n)},L_{m'}^{(n')}\Bigr]\ \tau
=\lag 0|e^{H(t)}\Bigl[\hat{\cal L}_{m}^{(n,0;+)},\hat{\cal L}_{m'}^{(n',0;+)}
\Bigr]\ g|0\rag.
\ee
This means that $\{L_{m}^{(n)}\}$
is nothing but the generators of \sdv\ ($\subset$\dvr)
as the action on $\tau(t,g)$.

Since the generator of \sdv\ is represented by using the vertex operators,
we can investigate the relation between \sdv\ and
$W_{1+\infty}$.
Let us set $\{W_{p}^{(k)};\ p\in{\bf Z}, k\in{\bf Z}_{>0}\}$ as generators of
 $W_{1+\infty}$.
It is known that the generators of $W_{1+\infty}$
can also be written by using the vertex operators \cite{dic2,tt};
\be
W^{(k)}(z)=\s_{p=-\infty}^{\infty} z^{-p-k}\ W_{p}^{(k)}=
\left. \biggl( {{\pa}\over{\pa \ze}} \biggr)^{k-1}
           Y(z,\ze)\ \right|_{\ze=z}.
\ee
Combining this equation with (\rf{shift}) and (\rf{tom}),
we verify the relation between $L_{m}^{(n)}$ and $W_{m}^{(k)}$ as
\bea
L_{m}^{(n)}\
&=& -\s_{k=1}^{\infty} {e^{{i\pi\over2}k} \over (k-1)!}\
{{\Bigl(2\sinh({\lambda n/2})\Bigr)^{k-1}
   \cosh\Bigl( \{\lambda n(m+k+1)-i\pi k\}/2\Bigr) }
\over{\sinh\lambda n}}\ W_{m}^{(k)}    \nonumber\\
&\equiv&\s_{k=1}^{\infty} (s_{m})_{nk}\ W_{m}^{(k)}
\lb{rel}
\eea
for fixed $m$.
It is also written in the matrix form as
\be
{\bf L}_{m}={\bf S}_{m}{\bf W}_{m},
\lb{mat}
\ee
where the matrices are defined by
\bd
{\bf L}_{m}={}^{T}(L_{m}^{(1)},L_{m}^{(2)},\cdots),\ \
{\bf W}_{m}={}^{T}(W_{m}^{(1)},W_{m}^{(2)},\cdots),\ \
{\bf S}_{m}=(s_{m})_{nk}\hspace{.2in}(n,k\in {\bf Z}_{>0}),
\ed
respectively.
Since we can consider the following correspondence as
\be
z^{m+k}\pa_{z}^{k}\ v\longleftrightarrow W_{m}^{(k)}\ \tau\ \ ;\ \
z^{m}\left[ z\pa_{z} +{m \over 2}  \right] _{n;-} v\longleftrightarrow
 L_{m}^{(n)}\ \tau,
\ee
we can say ${\bf S}_{m}$ has the inverse similar to $c_{jk}$ in (\rf{cjk})
(see (\rf{cv})).
Then (\rf{mat}) is also written the form as
${\bf W}_{m}={\bf S}_{m}^{-1}{\bf L}_{m}$.
Hence we conclude that $\{L_{m}^{(k)}\}$ is another realization of
$\{W_{m}^{(k)}\}$, that is \sdv\ is nothing but $W_{1+\infty}$.
Though we sum up the index $k$ in (\rf{rel}), i.e. we forget the
information on the conformal spins, we recover their dependence via the
index $n$ of $\{L_{m}^{(n)}\}$.
Mathematically, $n$ decides the power of $q$, in other words,
the interval of discretization.
But from the field theoretical point of view, we can say
the difference between \sdv\ and
$W_{1+\infty}$ owes to the difference of the corresponding currents
as follows:
{
\setcounter{enumi}{\value{equation}}
\addtocounter{enumi}{1}
\setcounter{equation}{0}
\renewcommand{\theequation}{\theenumi\alph{equation}}
\bea
\mbox{\sdv} &\ra&\ L^{(n)}(z)=\s_{m}z^{-m-1} \ L_{m}^{(n)}\\
W_{1+\infty}\ \ &\ra&\ W^{(k)}(z)=\s_{m}z^{-m-k}\ W_{m}^{(k)}.
\eea
\setcounter{equation}{\value{enumi}}
}

\section{Concluding remarks}

In this paper, we have found that the simplest \dv\ subalgebra, \sdv,  can
be regarded as the $W_{1+\infty}$ algebra.
The implication of this fact is very impressive:
The $W$-infinity symmetry
is considered as a universal symmetry structure for integrable systems
in a sense that such symmetry often appears in various theories.
On the other hand, as mentioned at the beginning, some discretization of the
variables preserve integrability of the system.
Therefore \sdv\ is not only a deformation of the \vi\ algebra
but a discretization of the spectral parameter which
preserves integrability of the KP hierarchy.
This point of view must provide useful tools
for understanding structure of integrable systems,
such as the soliton theories,
2D gravity, etc.

\vspace{.3cm}

The following is devoted to some remarks on the subject:

\noindent(i)\ As mentioned above, we have not used the whole structure of
\dvr.
The remaining part must also have rich structure:\\
\noindent$\circ$ If we consider $D$-Vir$^{(+,r\neq0)}$ algebra, the
fermion is no longer unique constituent.
In this case it is natural to think that the representation is also
``discretized'', i.e. we must consider deformation of the KP hierarchy
itself.
It is interesting whether such system is still integrable or not.\\
\noindent$\circ$ Another type of \dv\ subalgebra, $D$-Vir$^{(-)}$,
is also related with the $W$-infinity algebra.
{}From the fact that the generator of $D$-Vir$^{(-,r=1)}$ is realized by
ordinary free bosons in (\rf{bil}),
it is reasonable to guess that it is
 related to the $W_{\infty}$ algebra \cite{prs}.

\vspace{.2cm}

\noindent(ii)\ Though we made no mention of in this paper,
there exists structure of the Moyal bracket \cite{moy}
behind \dv\ \cite{kem}.
The structure is so huge that it contains not only \dv\ but also deformation
of some infinite dimensional Lie algebras \cite{prs,f}.
It suggests that the Moyal structure must have many informations
on integrability.
Therefore study of the symmetry associated with this structure
seems valuable to have general view of integrable systems.

\end{document}